# A versatile variable field module for field and angular dependent scanning probe microscopy measurements

Hongxue Liu [a], Ryan Comes [a], Jiwei Lu [a], Stuart Wolf [a,b], Jim Hodgson [c], Maarten Rutgers [c]

[a] Department of Materials Science and Engineering, University of Virginia, Charlottesville, VA 22904
[b] Department of Physics, University of Virginia, Charlottesville, Virginia 22904
[c] Asylum Research, Santa Barbara, CA 93117

We demonstrate a versatile variable field module (VFM) with capability of both field and angular dependent measurements up to 1800 Oe for scanning probe system. The magnetic field strength is changed by adjusting the distance between a rare earth magnet and the probe tip and is monitored in-situ by a built-in Hall sensor. Rotating the magnet allows the field vector to change from the horizontal to vertical direction and makes it possible to do angular dependent measurements. The capability of the VFM system is demonstrated by degaussing a floppy disk media with increasing magnetic field. Angular dependent measurements clearly show the evolution of magnetic domain structures, with a completely reversible magnetic force microscopy phase contrast observed when the magnetic field is rotated by 180°. A further demonstration of out-of-plane and in-plane magnetic switching of $CoFe_2O_4$ pillars in $CoFe_2O_4$-$BiFeO_3$ nanocomposites was presented and discussed.

## I. Introduction

With the rapid progress of nanotechnologies, atomic force microscopy (AFM) has become one of the most important techniques for imaging and measuring matter at the nanoscale.[1] Far beyond its original use to image the topography of surfaces, AFM has also been widely used to measure other characteristics such as electrical, ferroelectric, magnetic properties, and to perform different types of spectroscopy.[2-5]

Many materials themselves and/or their formed composites have interesting magnetic field dependent properties (for example, conductivity, dielectricity, ferroelectricity, magnetism). Studying these materials under magnetic field at the nanoscale is not only important for the understanding microscopic nature of their properties, but also essential for potential applications. One example is the multiferroic materials which show simultaneous presence of multiple ferroic orders such as ferroelectricity and ferromagnetism.[6] The interplay of ferroic orders is one of the key properties that open large potential applications in memory and logic devices that can be controlled both electrically and magnetically.[7] Thus understanding these interplays under magnetic field is very important.

Currently there are several commercial options available for field dependent scanning probe microscopy measurements.[8-11] One example is the VFM for Asylum MFP-3DTM AFM. For all the options designed as accessories to existing AFM systems, they only provide in-plane (IP) magnetic fields. The options of attaching AFM modules to existing magnetic measurement systems such as Quantum Design PPMS allow for scanning probe measurements under out-of-plane (OP) magnetic fields. In this paper, we describe and demonstrate a new design of VFM which provides the capability of both field and full angular dependent scanning probe microscopy measurements. While the current module is specially made to fit inside the Cypher AFM from Asylum Research, it is a universal design and may be retrofitted to any existing scanning probe system as long as some modifications are made to the geometry.

**II. System Design**

Figure 1(a) and 1(b) show the schematic drawing and actual picture of the VFM mounted inside a Cypher AFM, respectively. The module consists of several integral parts including a rotating shaft to which a strong rare earth magnet is attached, a micrometer with both X and Y adjustments, and a mounting stage to attach the rotating shaft to the micrometer. The basic principle of the design is to change the magnetic field strength of the scanning area under the probe tip by adjusting the distance between the probe tip and a permanent magnet. The magnetic field vector can be readily changed by rotating the magnet. In order to quantitatively determine the magnetic field, a sample post with a built-in hall sensor right under the probe tip is fabricated. The hall sensor is calibrated with known magnetic fields and the output voltage is linear with the applied magnetic field as shown in the insert of Fig. 2(a).

In order to reach high magnetic fields, the magnet needs be placed near the probe tip, as close as possible. This leads to some limitations and design compromise due to the way that AFM works which relies on laser reflection from the top surface of the cantilever to measure the deflection of the cantilever. The magnet, when placed too close to the tip, will readily block the laser reflection and thus interfere with the operation. In this case, the smaller the size of the magnet, the closer the magnet can be placed to the probe tip. On the other hand, too short a distance between tip and magnet limits the scan region to the edge of a sample which is probably not the best place for characterizing the sample. In this case, a larger size magnet is preferred since it can extend the field to a higher value and also has better field resolution along the length. For these reasons a 1/4" cubic neodymium magnet (Grade N52, K&J Magnetics) is chosen which gives magnetic fields up to 1800 Oe without affecting the operation of the Cypher AFM. It should be noted that the module still has its limitation of achievable maximum magnetic field. Further upgrade is planned to increase the magnetic field by enhancing the magnet with magnetic materials with high permeability.

The distance dependent magnetic field strength is shown in Fig. 2(a). As expected, the field follows an exponential-like decrease along the distance. However, due to the fact that the distance is comparable to the size of the magnet, the curve can not be fitted to $H \propto 1/d^3$ relationship. In order to overcome the limited adjustment range of the micrometer, there are two additional soft lock positions of the rotating shaft relative to the mounting stage along the X-direction to further extend the range of distance of the magnet from the probe tip. This produces magnetic fields from less than 100 Oe to 1800 Oe.

**III. Performance and Results**

To test the capability and performance of the VFM system, a series of field and angular dependent magnetic force microscopy (MFM) measurements were performed on a floppy disk media. For all the measurements, a beryllium copper non-magnetic cantilever clip was used to minimize field distortion and a high coercivity (> 5000 Oe) MFM probe was used. As shown in Fig. 2(b), the field dependent magnetic properties of the floppy disk media measured at 300 K show it has a strong IP magnetic anisotropy. Figure 3 shows the MFM images taken under different magnetic fields. The distinct blue and red stripe contrast in the MFM image taken without magnetic field clearly represents the magnetic bit profile. After applying increasing IP magnetic fields, the stripe contrast starts to disappear, indicating that the magnetic bit profile is progressively destroyed. The written bits are erased at around 850 Oe, which is close to the coercivity of 830 Oe determined from the M-H curve in Fig. 2. With further increasing fields up to 1750 Oe, no significant changes of phase contrast and roughness were observed, indicating the magnetization was quickly uniformly aligned. The degaussing process of the floppy disk media using OP magnetic fields was also tried. As shown in Fig. 4, the written bits are erased at a much slower rate with increasing magnetic fields. Because the bits are originally magnetized in the IP direction, higher energy is required to rotate the magnetization from this preferred direction to out of plane. At 1750 Oe, the stripe contrast is faint but still distinguishable.

Figure 5 shows the angular dependent MFM images taken under 1750 Oe starting from an IP magnetic field at 0 degree, clearly demonstrating the evolution of magnetic domain structures under fields with different rotation angles relative to the original IP fields used for degaussing. A completely reversible MFM phase contrast is observed when the magnetic field is rotated by 180°.

The capability of angular dependent measurements was further tested by trying to switching the $CoFe_2O_4$ magnetic pillars in $CoFe_2O_4$-$BiFeO_3$ (CFO-BFO) composites. Fig. 6 (a, c) and (b, d) show the MFM images of a CFO-BFO composite film grown on a $SrTiO_3$ (001) substrate co-deposited by pulsed electron deposition (PED),[12] after applying magnetic fields in the OP and IP directions, respectively. While a uniform OP or IP magnetization can not be explicitly determined and perfectly matched to the height profile of CFO pillars, some CFO pillars show clear signs of switching to either OP or IP dependent on the applied field vector directions, as shown in the reversible blue and red contrast marked in the circle areas. The difference of MFM images under OP and IP magnetic fields is also clearly observed. The absence of uniform magnetization along the applied field can probably be ascribed to the following reasons. First the pillar density (pillar distance), size, and aspect ratio in a co-deposited sample do not have a perfect uniformity, which means microscopically not all the CFO pillars have the same magnetic characteristics (coercivity, anisotropy) measured macroscopically by a vibrating sample magnetometer. Some of the pillars with larger coercivities need higher magnetic fields to switch. Second the magnetic exchange interaction between very close CFO pillars can be strong and complicate things further.

## IV. Conclusions

We developed a versatile VFM with the capability of both field and angular dependent measurements up to 1800 Oe for scanning probe system. The magnetic field is changed by adjusting the distance between the probe tip and a rare earth magnet, and the magnetic field vector is changed by rotating the magnet. The field dependence and angular dependence measurement capabilities of the VFM system were successfully demonstrated by degaussing a floppy disk media and by studying the magnetic switching in a CFO-BFO nanocomposite.

**Acknowledgements**

The work at University of Virginia was supported by DARPA under contract no. HR-0011-10-1-0072 and the Nano-electronics Research Initiative through NSF under contract no. DMR-08-19762. Ryan Comes also wishes to acknowledge support from the National Defense Science and Engineering Graduate Fellowship.

**Figure Captions**

Fig. 1. The (a) schematic drawing and (b) actual picture of the VFM installed in a Cypher AFM system.

Fig. 2. (a) The distance dependent magnetic field measured by the Hall sensor. The insert shows the output Hall voltages calibrated with known magnetic fields. (b) The field dependent magnetization of a floppy disk media measured at 300 K.

Fig. 3. MFM images of a floppy disk media taken (a) before applying magnetic field, and after applying IP fields of (b) 500 Oe, (c) 800 Oe, (d) 850 Oe, (e) 900 Oe, and (f) 1750 Oe.

Fig. 4. MFM images of a floppy disk media taken (a) before applying magnetic field, and after applying OP fields of (b) 500 Oe, (c) 800 Oe, (d) 900 Oe, (e) 1200 Oe, and (f) 1750 Oe.

Fig. 5. Angular dependent MFM images of a floppy disk media taken at 1750 Oe starting from an IP magnetic field at 0 degree.

Fig. 6. MFM images of a CFO-BFO composite film taken at (a, c) OP magnetic field, (b, d) IP magnetic field. The applied magnetic field is 1750 Oe.

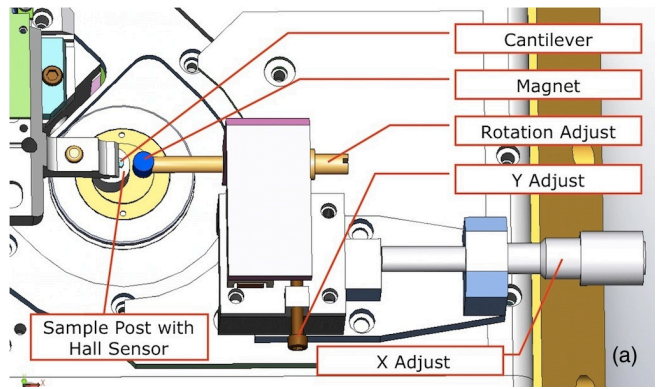

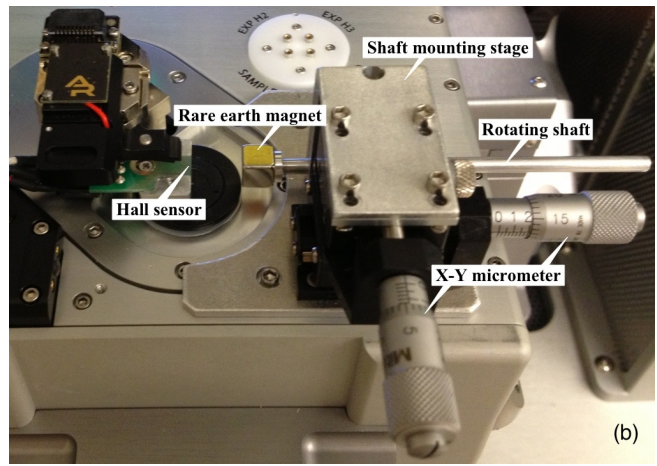

Fig. 1. Liu, et al.

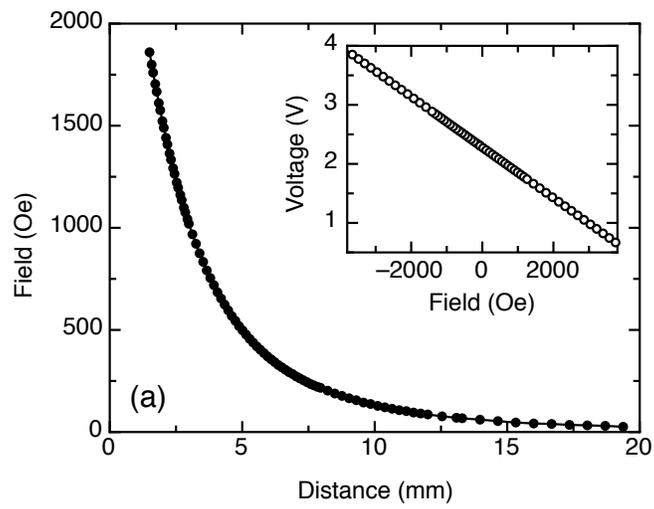

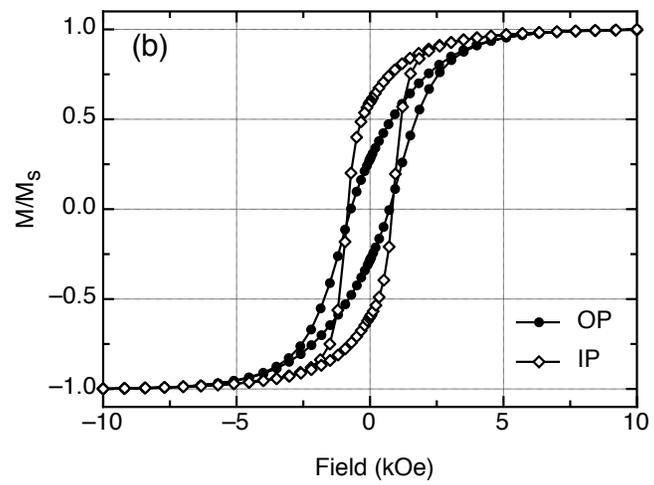

Fig. 2. Liu, et al.

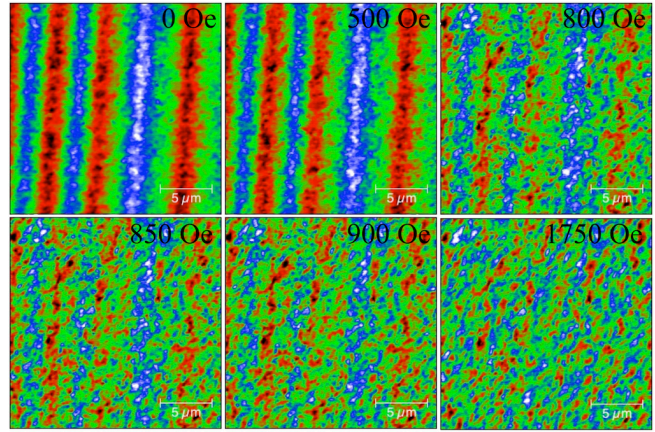

Fig. 3. Liu, et al.

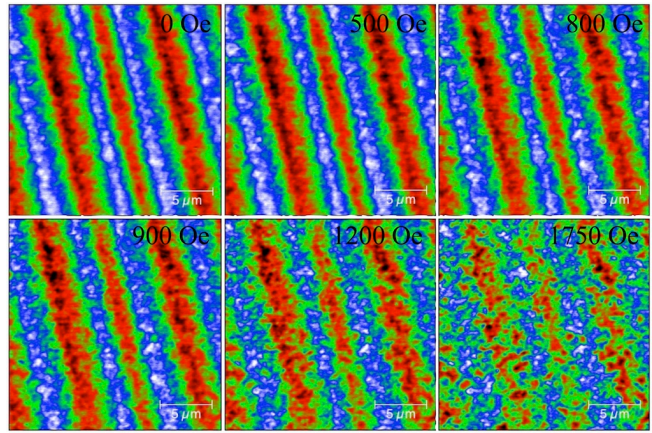

Fig. 4. Liu, et al.

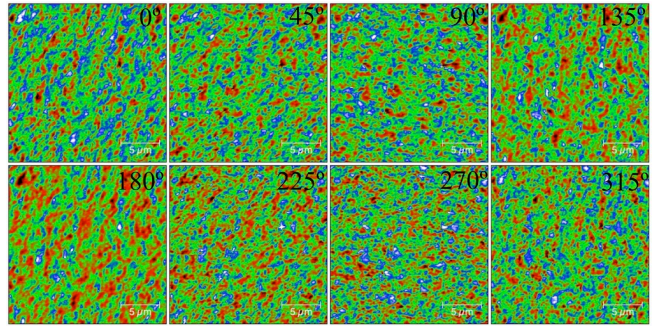

Fig. 5. Liu, et al.

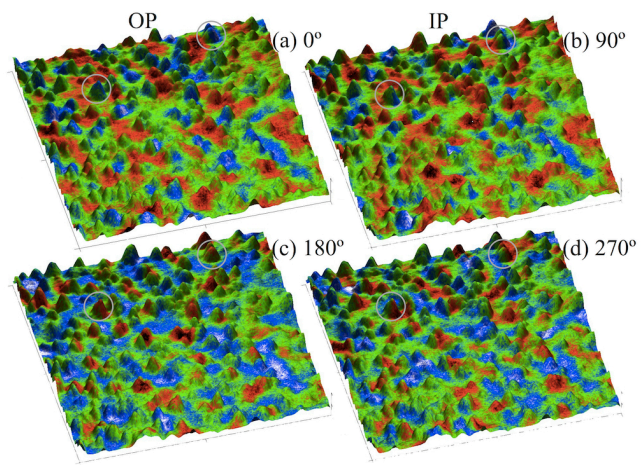

Fig. 6. Liu, et al.